\documentclass[prl,twocolumn,amsmath,amssymb,groupedaddress,superscriptaddress]{revtex4}
\usepackage{graphics}

\def\be{\begin{equation}}

\def\ee{\end{equation}}

\def\eeff{\textmd{eff}}

\def\ccol{\textmd{coll}}

\def\iint{\textmd{int}}

\def\be{\begin{equation}}
  \def\ee{\end{equation}}
\def\bea{\begin{eqnarray}}
  \def\eea{\end{eqnarray}}

\begin{document}

\title{Matter--wave emission in optical lattices: Single particle and collective effects}

\author{In{\'e}s \surname{de Vega}}
\email{Ines.devega@mpq.mpg.de}
\affiliation{Max-Planck-Institut
f\"ur Quantenoptik,
  Hans-Kopfermann-Str. 1, Garching, D-85748, Germany.}
\author{Diego \surname{Porras}}
\email{Diego.Porras@mpq.mpg.de}
\affiliation{Max-Planck-Institut
f\"ur Quantenoptik,
  Hans-Kopfermann-Str. 1, Garching, D-85748, Germany.}
\author{J. Ignacio \surname{Cirac}}
\email{Ignacio.cirac@mpq.mpg.de}
\affiliation{Max-Planck-Institut
f\"ur Quantenoptik,
  Hans-Kopfermann-Str. 1, Garching, D-85748, Germany.}

\begin{abstract}
We introduce a simple set--up corresponding to the matter-wave
analogue of impurity atoms embedded in an infinite photonic
crystal and interacting with the radiation field. Atoms in a given
internal level are trapped in an optical lattice, and play the
role of the impurities. Atoms in an untrapped level play the role
of the radiation field. The interaction is mediated by means of
lasers that couple those levels. By tuning the lasers parameters,
it is possible to drive the system through different regimes, and
observe phenomena like matter wave superradiance, non-Markovian
atom emission, and the appearance of bound atomic states.
\end{abstract}

\date{\today}

\maketitle
Recent progress in atomic physics has allowed experimentalists to
trap atoms in optical potentials at very low temperatures. This
has led to the observation of several interesting phenomena in
which atom--atom interactions play a predominant role. With atoms
loaded in optical lattices it is nowadays possible, for example,
to reach the strong correlation regime where quantum phase
transitions between superfluid and insulator phases \cite{J98,G02},
Tonks--Girardeau gases \cite{PC04}, or even entanglement between neighboring
atoms can be observed. Those experiments have triggered a large
amount of theoretical work proposing and analyzing new experiments
where intriguing condensed matter behavior could be observed.

\begin{figure}
  \resizebox{\linewidth}{!}{%
    \includegraphics{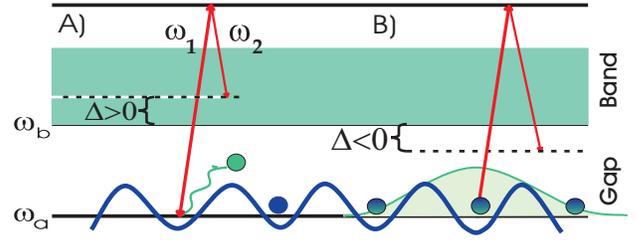}}
    \caption{(Color online) Schematic diagram: Lasers produce
    two-photon Raman transitions to an untrapped state. Fig A: For
    laser detunings $\Delta>0$, transitions are into a \textit{band} of
    non--trapped states. Fig B: $\Delta<0$, transitions are into the
    \textit{gap} region, and a trapped-untrapped atom bound state is
    formed.} \label{fig0}
\end{figure}

In this work we show that with the same systems it is possible to
observe a broad spectrum of different phenomena usually connected
to light--matter interactions (see also \cite{RFZDZ04,PPO07,M02}
for related setups). In our setup, the role of matter is played by
the absence/presence of one atom in the ground state of an optical
potential, whereas the role of light is played by
weakly--interacting atoms in a different internal state which is
not affected by the optical potential. The coupling between those
two systems is induced by Raman lasers, which connect the two
internal states of each atom (see Fig. \ref{fig0}). As we will
show, the Hamiltonian that describes this situation is very
similar to that describing the interaction between two--level
atoms and the electromagnetic field within a photonic crystal
(PC). By changing the laser and optical trapping parameters it is
possible to drive the system to different regimes where a rich
variety of phenomena can be observed. These include the
spontaneous polarization of the system predicted by the mean field
theory \cite{JQ95}, collective effects in the emission of atoms
from the lattice \cite{Dicke,rusos,WYE07}, and the formation of a
bound trapped--untrapped atom state, analogous to the atom--photon
bound state that appear when atoms within a photonic crystal emit
photons within the gap region \cite{Joh87,JQ94,WJ03,DVAG05}.
Moreover, it is possible to reach a regime in which weakly
confined atoms drive atom--atom interactions between strongly
confined ones, giving rise to effective Coulomb-like interactions
between them.

We consider $N$ cold atoms with a ground state hyperfine level
$|a\rangle$ and frequency $\omega_a^0$ that is trapped by an
optical lattice with $M$ sites and lattice period $d_0$. The
motion of the atoms is restricted to the lowest Bloch band in the
collisional blockade regime, where either one or no atom occupy
each potential well of the lattice, which we will approximate by a
harmonic oscillator of frequency $\omega_0$. Then, we can replace
the creation operator at each site by $\sigma^+_{\bf
j}=|1\rangle_{\bf j} \langle 0|$, which describes transitions from
the Fock state $|0\rangle_{\bf j}$ with no atoms at site ${\bf
j}$, to a state $|1\rangle_{\bf j}$ with one atom \cite{RFZDZ04}.
The atoms have an additional internal level, $|b\rangle$, that is
not affected by the lattice potential (what can be achieved with
state dependent potentials) and has a frequency $\omega_b$. We
introduce the field operator $\psi^\dagger_b ({\bf
r})=(1/\sqrt{V})\sum_{\bf k}b^\dagger_{\bf k}e^{i {\bf k}\cdot{\bf
r}}$, where $V$ is the quantization volume, and $b^\dagger_{\bf
k}$ is the creation operator of an atom in $|b\rangle$ with
momentum ${\bf k}$.

In a similar way as in an atom laser setup \cite{MS97}, two lasers
are then used to induce two--photon Raman transitions between $|a\rangle$ and $|b\rangle$. The lasers have a two--photon Rabi
frequency $\Omega$, and their frequencies and momentum differences
are $\omega_L=\omega_{1}-\omega_2$ and ${\bf k}_L={\bf k}_1-{\bf
k}_2$ respectively. When tuning them close to a two photon
resonance and far from single photon resonances, an effective
Hamiltonian is obtained which in the interaction picture can be
written as ($\hbar=1$)
\begin{eqnarray}
H_{\iint}&=&\sum_{\bf j} \sum_{\bf k}g_{k} \left(b^\dagger_{\bf
k}\sigma_{\bf j} e^{i \Delta_{\bf k} t-i({\bf k}-{\bf k}_L)\cdot{\bf
r}_{\bf j}}+h.c.\right). \label{genH2}
\end{eqnarray}
Here ${\bf r}_{\bf j}$ denotes the positions in the lattice, and
$\Delta_{\bf k}=k^2/2m -\Delta$, with $\Delta=\omega_L-(\omega_b
-\omega_a)$ the laser detuning (and
$\omega_a=\omega_a^0+\omega_0/2$). The coupling constants are $g_k
= \Omega e^{-X_0^2 k^2/2} (8\pi^{3/2} X_0^3/V)^{1/2}$, where
$X_0=(1/2m\omega_0)^{1/2}$ is the size of the wave function at each
site.

The similarity of Hamiltonian (\ref{genH2}) with that describing
the interaction of atoms with the electromagnetic field is
apparent. The dispersion relation of the atomic bath is contained
in $\Delta_k$, and resembles that of the radiation field in a
three dimensional and infinite PC near the band-edge
\cite{JQ94,JQ95,DVAG05}. Furthermore, in our set--up one can
easily control several external parameters: $\Omega$, which
determines the coupling strength, $\Delta$, which determines the
resonance conditions, the number of atoms $N$ and of sites $M$, the lasers wavevectors ${\bf k}_L$, and the dimension of the trap and the lattice. Thus, we expect to observe a rich variety of phenomena with our system,
ranging from some well--known from the field of Quantum Optics, to
other which are difficult to access in that field.
We will
start out with a single excitation ($N=M=1$), and then
consider collective effects ($M>1$), for the different regimes
dictated by the control parameters $\Omega$ and $\Delta$.

\begin{figure}
  \resizebox{\linewidth}{!}{%
    \includegraphics{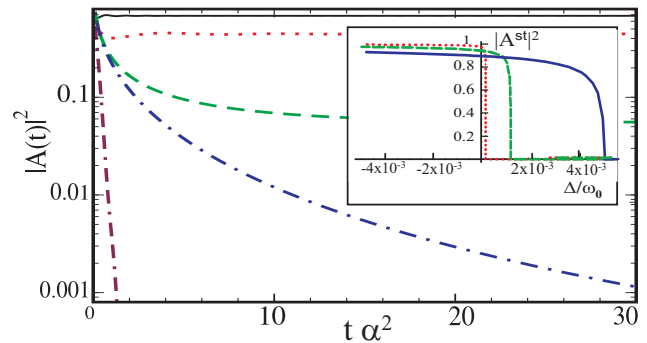}}
     \caption{Evolution of the atomic population $|A(t)|^2$ in logarithmic scale for different detunings. Solid, dotted, dashed, dot-dashed and dot-dot-dashed lines correspond respectively to $\Delta/\alpha^2=-8,-1,-0.2,0.2,8$ ($\omega_0=\infty$), where one can recognize the regimes explained in the text. Inset \cite{general}: Steady state population $|A^{st}|^2=|A(\infty)|^2$. Solid, dashed and dotted lines correspond respectively to  $\Omega/\omega_0=0.05,0.025,0.01$.} \label{exact}
\end{figure}

An atom within a single trap constitutes the simplest setup but it
still gives a very good insight into the problem. The wave
function of the system has the form
$|\Psi(t)\rangle=A(t)|1,\{0\}\rangle+\sum_{\bf k}B_{\bf
k}(t)|0,{1}_{\bf k}\rangle$, where $|1,\{0\}\rangle$ describes the
atom in the trapped state and no free atom present, and
$|0,{1}_{\bf k}\rangle$ represents no atom in the trapped state
and a single untrapped atom in the mode ${\bf k}$. Using the
Schr{\"o}dinger equation we have $\dot A(t)=-\int_0^t d\tau
G(t-\tau)A(\tau)$, where
\be
 G(t)=\sum_{\bf k} g_k^2 e^{-i\Delta_{\bf
k}t}=
 \Omega^2\frac{e^{i (\Delta t-\arctan{[\omega_0 t])}}}{\nu_t |\nu_t|^2},
 \label{Gn0}
 \ee
with $\nu_t=2 \sqrt{1+i\omega_0 t}$,
is the correlation function of the environment.
An analytical solution can be obtained by assuming that trapped atoms are strongly confined so that $\omega_0 \gg \Omega,\Delta$.
For a $3D$ bosonic field, this leads to $G_\infty(t)=-\alpha e^{i(\Delta
t+\pi/4)}/t^{3/2}$, which is singular at the origin, but describes
correctly times $t\gg 1/\omega_0$. Except for the value of $\alpha=\Omega^2/\omega_0^{3/2}$,
 $G_{\infty}(t)$ is identical to the
correlation function of the radiation field within an anisotropic
PC, as described in \cite{WJ03}. In the same way, a $1D$
environment (produced by trapping the atoms in
$|b\rangle$ in a $1D$ harmonic trap), gives rise to a
correlation function similar to the one appearing for the
radiation field in isotropic PCs.
Using the Laplace transform method, we get \cite{WJ03}
\begin{equation}
A(t)=c_1 e^{i (r_1^2 +\Delta)t}+I(\alpha,\Delta,t),\label{sol}
\end{equation}
with $I(\alpha,\Delta,t)=(\alpha e^{i\pi/4}/\pi) \int_0^\infty dx
\frac{\sqrt{x}e^{(-x+i \Delta)t} }{(-x+i \Delta)^2+i \alpha^2 x}$. Defining $r_{\pm}=-(\alpha/2)\pm\sqrt{(\alpha/2)^2-\Delta}$ we have: (i) If $\alpha^2/2> \Delta >0$, then $c_1=0$. (ii) If
$\Delta>\alpha^2/2$, then $r_1=r_-$ and $c_{1}=\frac{2 r_-}{r_--r_+}$. (iii) If $\Delta <0$, $r_1=r_+$ and $c_1=\frac{2r_+}{r_+ -r_-}$. Depending on the parameters, we can have very
different behaviors. For $\Delta >0$ there is no trapped atom left
in steady state, whereas for $\Delta<0$ this
is not the case (Fig. \ref{exact}). Thus, there is a quantum
phase transition at $\Delta=0$ analog to
that found in the spin--boson model \cite{spinboson}. Moreover,
for $\Delta<0$, the emitted atomic field is in the form of evanescent modes exponentially
localized around the trapped atom, what leads to the trapped
atom--untrapped atom bound state. This follows from the
probability of finding a radiated particle at position ${\bf r}$
at a long time $t$, $|\Psi_b({\bf r},t)|^2=(c_1 \Omega m X^{3}_0
/2\pi r)^2 e^{-\textmd{Im}[k^e_0] r}e^{-\textmd{Im}[r_1^2] t}$,
where $k^e_0=\sqrt{2m(\Delta-r_1^2)}$ is imaginary if $\Delta<0$.
For $\Delta>0$ we have two different regimes: for $\Delta\gg
\alpha^2$ we just have an exponential rate $\Gamma_0=2\Omega^2\sqrt{\pi\Delta/\omega_0^3}$,
whereas in the opposite limit the evolution does not follow such a
law. The first corresponds to the Markovian regime, where the
correlation time $\tau_c\simeq \Delta^{-1}$ of the environment (here
untrapped atoms) is shorter than the typical evolution time of the
trapped atom ($\Gamma_0^{-1}$) \cite{general}.



We now study the dynamics of atoms in a lattice with $M$ sites, choosing $\omega_0\gg\Omega,\Delta$.
Guided by the previous analysis, we will consider the regimes
$\Delta>0$ (Markovian and non--Markovian), as well as $\Delta<0$.


In the limit where $\Gamma_{\rm coll} \tau_c\ll 1$, where
$\Gamma_{\rm coll}$ gives the typical evolution time of the trapped
atoms, we can analyze the problem under the Born--Markov
approximation. The dynamics of the atoms in the lattice is
dictated by the quantities
\begin{eqnarray}
\Gamma_{|{\bf i-j}|}=\int_0^\infty d\tau G_{\bf i-j}(\tau )=
i|\Gamma_0|\xi \frac{e^{-\nu|{\bf i-j}|/\xi+i{\bf r_{i-j}}\cdot{\bf k}_L}}{|{\bf i-j}|},
\label{rate}
\end{eqnarray}
for ${\bf i}\neq {\bf j}$.
Here, $\nu=i,1$ for $\Delta>0$ and
$\Delta<0$, respectively, and the correlation function $G_{{\bf i-j}} (t)=\sum_{\bf k}g^2_{k}
 e^{i{\bf r}_{\bf i-j}\cdot({\bf k}-{\bf k}_L)-i\Delta_k t }$ is now
\begin{equation}
G_{{\bf i-j}} (t)=G(t)e^{ir_{\bf i-j}^2´/(4 X^2_0 \nu_t^2)}
 \label{Gn}
\end{equation}
with $G(t)$ given by (\ref{Gn0}).
Similar to the radiative case, the coefficients $\Gamma_{|{\bf
i-j}|}$ describe the dipolar interactions between the sites ${\bf
i}$ and ${\bf j}$. The quantity $\xi =1/(|k_0|d_0)$, with $k_0=\sqrt{2m\Delta}$ \cite{k0}, quantifies the range of
the interactions which, according to (\ref{rate}), has a Yukawa form.

For $\Delta>0$, the situation under study resembles that of a set
of $M$ atoms in a lattice of constant $d_0$ interacting with the
electromagnetic field with resonant wave--vector $k_0$, and where
$N$ corresponds to the number of excited atoms. Thus, phenomena
like multiple--scattering, reabsorption, or superradiance should
be expected. In fact, apart from $\xi$ we can also define
the analogous of the optical depth, which for a cubic lattice is
$\chi=M^{1/3}\xi^2$. Depending on the values of those
dimensionless parameters we can predict different phenomena.

Let us first consider the case of one atom $N=1$, symmetrically
distributed through a lattice of $M$ sites. We will assume that
laser directions are chosen such that $|{\bf k}_L|=|k_0|$. The
state at time $t$ can be expressed as
$|\Psi(t)\rangle=(1/\sqrt{M})\sum_{\bf i} A_{\bf i}(t)|1_{\bf
i},\{0\}\rangle+\sum_{\bf k}B_{\bf k}(t)|0,{1}_{\bf k}\rangle$,
where $A_{\bf i}$ represents the amplitude at site ${\bf i}$.
Following similar lines as in the single site example we can write
$\dot A_{\bf i}(t)=\sum_{\bf j}\Gamma_{|{\bf i-j}|}A_{\bf j}(t)$,
with rates given by (\ref{rate}). An analytical solution can be
obtained by considering a large system ($M^{1/3}\gg 1$) such that
boundary effects are neglected, and periodic boundary conditions
can be assumed. In that situation, atoms in $|a\rangle$ remain in
the completely symmetric state with an amplitude
$A_{\textmd{coll}}(t)=(1/\sqrt{M})\sum_{\bf i}A_{\bf i}(t)$ that
decays with a collective rate
$\Gamma_{\textmd{coll}}=\sum_{\bf{n}}\Gamma_{|{\bf n}|}$.  Then,
the system reaches several regimes in which collective effects
play an important role \cite{Diego}: (a) If $\xi<1$ and $\chi\gg
1$, reabsorption occurs, and the decay rate is renormalized to
$\Gamma_{\textmd{coll}}\sim \chi\Gamma_0$. (b) If $M^{1/3}>\xi>1$,
the dipolar interactions couple near neighbors, and $\chi>1$
independent of the size of the system. The rate
$\Gamma_{\textmd{coll}}$ scales in the same way as in (a).
Finally, case (c) corresponds to $\xi\gg M^{1/3}$, a situation in
which every site is connected through dipole interactions to all
other sites, and $\Gamma_{\textmd{coll}}= M\Gamma_0$.
%

\begin{figure}
  \resizebox{\linewidth}{!}{%
    \includegraphics{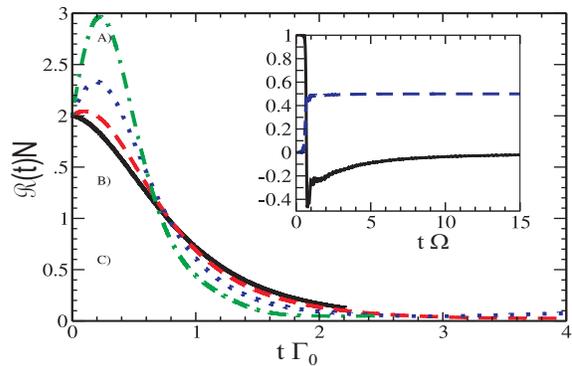}}
    \caption{Rate of atomic emission from a $1D$ lattice with
    $N=M=100$ atoms. 
    Solid, dashed, dotted and dot-dashed curves correspond to $\xi=0.9,1.25,2,3.33$ respectively.
    Inset \cite{general}: Spontaneous symmetry breaking and non--zero
    steady state population. Solid and dashed curves represent $z(t)$ and $y(t)$ respectively, evolving according to
    (\ref{semic}), for $N=M=10^3$ atoms, $\Delta=0$, and $\omega_0=50$.}
 \label{super}
\end{figure}

We now consider $N$ atoms within $M$ sites. The above described
regimes still give a valid picture for this situation. In
particular, we focus on the collective limit (b), where
$\xi>1$. In addition, from here on, we will consider laser
directions to be such that $|{\bf k}_L|d_0 M^{1/3}\ll 1$. When all the atoms are initially in the lattice, we get
\begin{eqnarray}
 \frac{d\langle\sigma^{3}_{\bf i} \rangle}{dt}&=&-4 {\textmd Re}
 [\sum_{\bf j} \Gamma_{|{\bf j-i}|}\langle \sigma^+_{\bf i}
 \sigma_{\bf j} \rangle]\nonumber\\
 \frac{d\langle \sigma^+_{\bf i} \sigma_{\bf j} \rangle}{dt}&=&
 \sum_{\bf l} \Gamma^*_{|{\bf l-i}|}\langle \sigma^+_{\bf l}
 \sigma^3_{\bf i} \sigma_{\bf j} \rangle+\Gamma_{|{\bf l-j}|}\langle
 \sigma^+_{\bf i} \sigma^3_{\bf j} \sigma_{\bf l} \rangle,
 \label{sistem0}
\end{eqnarray}
with rates given by (\ref{rate}). Here, $\sigma_{\bf
i}^{3}=2\sigma^+_{\bf i} \sigma_{\bf i}-1$, and all the
operators are evaluated a time $t$. Let us first analyze the
atomic emission that occurs for positive detuning $\Delta>0$. We
focus on the rate of emission of atoms in all directions, which is
given by ${\mathcal R}(t)\approx -\sum_{\bf j}d\langle
\sigma^3_{\bf j}\rangle/dt$, for different values of  $\xi$. If sites evolve independently, $\mathcal{R}(t)$ decays exponentially. However, when $\xi>1$ and
collective effects are present, ${\mathcal R}(t)$ does no longer
decay exponentially and, furthermore, it presents positive slopes
at initial times. This is shown in Fig. \ref{super} for a $1D$ lattice, where it is observed that collective effects occur for $\xi>1$.
This result is obtained with (\ref{sistem0}) by using the semiclassical
decoupling $\langle \sigma^+_{\bf l}  \sigma^3_{\bf i} \sigma_{\bf
j} \rangle=\langle \sigma^3_{\bf i}\rangle\langle \sigma_{\bf l}^+
\sigma_{\bf j} \rangle$, that is based on neglecting atomic
quantum fluctuations \cite{rusos}. Nevertheless, the change of sign in the slope can be obtained analytically by differentiating Eq.
(\ref{sistem0}) at $t=0$ without the use of any approximation.

If we now consider negative detuning, $\Delta<0$, the rates
(\ref{rate}) are purely imaginary, and the system has an effective
Hamiltonian
\begin{eqnarray}
H^{\Delta<0}_{\eeff}=\sum_{\bf i,j} J_{|{\bf i}-{\bf j}|}\sigma_{\bf i}^\dagger \sigma^-_{\bf j} ,
\label{Hoping}
\end{eqnarray}
where $J_{|{\bf i}-{\bf j}|}=i\Gamma_{|{\bf i}-{\bf j}|}$ is a
real and negative quantity, so that (\ref{Hoping}) is Hermitian
and describes a coherent spin--spin interaction of ferromagnetic
type. This interaction may have interesting applications in the
field of quantum simulation. Furthermore, for $\xi\gg 1$ it gives a
Coulomb--like interaction very difficult to obtain with
other techniques.

%

We now concentrate in the non--Markovian limit, where the system
also becomes strongly interacting. 
We will consider that all the atoms are initially in the lattice (i.e., $N=M$), and the limit
 $M\gg 1$. We use the mean field or Hartree
approximation \cite{Breuer}. Then, the evolution of
$y(t)=\sum_{\bf j} \langle\sigma^-_{\bf j} (t)\rangle/M$ and
$z(t)=\sum_{\bf j} \langle\sigma^3_{\bf j} (t)\rangle/M$ can be
written as
\begin{eqnarray}
\frac{dy(t)}{dt}&=&M\int_0^t d\tau G_{\ccol} (t-\tau) y(\tau)z(t);\nonumber\\
\frac{dz(t)}{dt}&=&-4 M \textmd{Re}\left[\int_0^t d\tau G_{\ccol} (t-\tau)y^* (\tau)y(t)\right].
\label{semic}
\end{eqnarray}
The function $G_{\ccol}(t)=\sum_{\bf n} G_{\bf n}(t)$, with
$G_{\bf n}(t)$ defined in (\ref{Gn}). Due to the non--Markovian
structure of the equations, the mean field approximation here
considered predicts that the trapped atoms acquire a macroscopic
polarization in the steady state. This is shown in Fig.
(\ref{super}) for $\Delta=0$, where we have considered an initial
infinitesimal polarization, $y(0)=10^{-6}$ (though the steady
state does not depend on the choice of $y(0)$), and $z(0)=1$. The
spontaneous polarization of the system, previously described in
\cite{JQ95} for atoms in PCs, is similar to the spontaneous
symmetry breaking described in the semiclassical theory of the
laser (\cite{Breuer} and references therein). Fig. (\ref{super})
also shows that the non--Markovian effects lead to a non--zero
steady state population, i.e. $z^{st}\neq -1$.
%
%

Most of the phenomena described here may be observed with state of
the art experimental setups using state--dependent potentials
\cite{J98} in the Mott insulator regime
\cite{G02} for the lattice atoms and choosing $\Omega\ll
\omega_0$ to avoid the occupation of other bands. The simplest
regime corresponds to the a single excitation ($N=M=1$), where
decay, non--Markovian effects, as well as the phase transition
occurring at $\Delta=0$ can be observed. Note that it is not required to have a
single atom in the whole lattice, as long as the atoms do not
interact with each others (i.e. $\xi\ll 1$ and initially
localized), so that one could easily monitor the decay as a
function of time by simply measuring how many atoms remain in $|a\rangle$. The presence of the bound state of the untrapped atoms
should be visible in the momentum distribution after free
expansion.
By preparing a few atoms each of them coherently distributed among $M$ sites in disjoint regions, several copies of the setup consisting in $N=1$ atom within $M$ sites could be realized. Hence, we would observe collective effects in the decay time for $\chi\ge 1$.
For an initial Mott insulator state
($N=M$) it should also be possible to observe superradiant effects
by looking at the slope of the decay rate for short times, and for $\Delta<0$ the
nearest--neighbor interaction induced via virtual transitions to
the untrapped state, as follows from Eq.\ (\ref{Hoping}). All those
phenomena require $\xi\ge 1$, i.e. $|\Delta|\le 1/(2md_0^2)$, as
well as the Markovian limit, $\Gamma_{\textmd{coll}}\le |\Delta|$. Observing
superradiance for long times (i.e. the whole shape of Fig.
\ref{super}) may be limited by decoherence effects caused by
random magnetic fields which shift $|a\rangle$ and $|b\rangle$ differently.
 Furthermore, to observe coherent interactions in Eq.\ (\ref{Hoping}) beyond nearest neighbors requires $\xi\gg 1$, which may also be compromised by decoherence
effects. A possible way around this is to use lighter atomic
species, like Li, where those conditions are relaxed. On the other
hand, the collecive non--Markovian effects related to the
spontaneous polarization should be also easy to observe by
choosing $\Delta\simeq 0$.

The proposed set--up may
be advantageous to observe some phenomena with respect to 
atoms interacting with light in a PC. First, it is
easily tunable; second, the detection techniques developed for atoms in optical lattices
\cite{Bloch} may allow to measure
features, like the
analogue to the
photon-atom bound state, that are difficult to measure in PCs;
third, the optical lattice is a nearly perfect
periodic potential. In addition, other interesting phenomena could be explored with
the present set--up.
For example, for $\chi\gg 1$ light--matter interface schemes can be used to control
the emission direction of the atoms, or to map the state of
trapped atoms into that of untrapped atoms \cite{Polzik}. Besides
that, if $|b\rangle$ is affected by a trap that is wider
than that of $|a\rangle$, other types of interactions, like the
one described by the Jaynes Cummings or the so-called Tavis
Cummings model, can be implemented. Finally, the present system
may be tuned to explore other regimes which have never been considered in
quantum optics since they could not be reached there, like
for example the ones in which the initial state of the atoms in
the lattice is a superfluid, or a Tonks gas.

We thank M. Aguado, M.C. Ba{\~n}uls, S. D{\"u}rr and G. Giedke for fruitful
discussions. Work supported by EU projects (SCALA, COMPASS), and DFG Munich-Centre for Advanced
Photonics. I.D.V acknowledges support from Ministerio de
Educaci\'on y Ciencia.


\begin{thebibliography}{99}
\bibitem{k0} Note that $k_0^e\approx k_0$, within the weak coupling limit.
\bibitem{general} For finite $\omega_0$, $\Gamma_0=\int_0^\infty dt G(t)=\Omega^2(-2 i/\omega_0+2\sqrt{\pi\Delta/\omega_0^3}e^{-\Delta/\omega_0}(1+\textmd{Erf}[i\sqrt{\Delta/\omega_0}]))$, where the first term is the analogue to the Lamb shift. In this case, the structure of $A(t)$ is similar to $(\ref{sol})$, but the phase transition above described is shifted to $\Delta=2\Omega^2/\omega_0$ (see inset Fig. \ref{exact}).
\end{thebibliography}
\end{document}